\documentstyle[12pt]{article}
\def\1ad{\mbox{\normalsize $^1$}}
\def\2ad{\mbox{\normalsize $^2$}}
\def\3ad{\mbox{\normalsize $^3$}}
\def\4ad{\mbox{\normalsize $^4$}}
\def\5ad{\mbox{\normalsize $^5$}}
\def\6ad{\mbox{\normalsize $^6$}}
\def\7ad{\mbox{\normalsize $^7$}}
\def\8ad{\mbox{\normalsize $^8$}}
\def\makefront{\vspace*{1cm}\begin{center}
\def\newtitleline{\\ \vskip 5pt}
{\Large\bf\titleline}\\
\vskip 1truecm
{\large\bf\authors}\\
\vskip 5truemm
\addresses
\end{center}
\vskip 1truecm
{\bf Abstract:}
\abstracttext
\vskip 1truecm}
\newcommand{\eqn}[1]{(\ref{#1})}
\newcommand{\ft}[2]{{\textstyle\frac{#1}{#2}}}
\def\be{\begin{equation}}
\def\ee{\end{equation}}
\def\bea{\begin{eqnarray}}
\def\eea{\end{eqnarray}}

\renewcommand{\a}{\alpha}
\renewcommand{\b}{\beta}

\renewcommand{\d}{\delta}
\newcommand{\pa}{\partial}
\newcommand{\g}{\gamma}
\newcommand{\G}{\Gamma}
\newcommand{\e}{\epsilon}

\renewcommand{\l}{\lambda}
\newcommand{\m}{\mu}
\newcommand{\n}{\nu}

\newcommand{\s}{\sigma}

\renewcommand{\o}{\omega}
\newcommand{\q}{\theta}

\newcommand{\del}{\partial}

\def\IZ{{\hbox{{\rm Z}\kern-.4em\hbox{\rm Z}}}}

\def\bigone{{\hbox{1\kern -.23em{\rm l}}}}

\newcommand{\NPB}[3]{{\it Nucl.\ Phys.} {\bf B#1} (#2) #3}

\newcommand{\PRD}[3]{{\it Phys.\ Rev.}\ {\bf D#1} (#2) #3}
\newcommand{\PLB}[3]{{\it Phys.\ Lett.}\ {\bf B#1} (#2) #3}
\newcommand{\tmath}[1]{\mbox{$#1$}}

\newcommand{\refer}[1]{(\ref{#1})}

\newcommand{\qb}{\bar{\q}}

\begin {document}
\begin{flushright} THU-98/06\\ {\tt hep-th/9802073}
\end{flushright}

\def\titleline{Supermembranes and Super Matrix Theory}
%%%%%%%%%%%%%%%%%%%%%%%%%%%%%%%%%%%%%%%%%%%%%%
%                                            %
% Insert now the text of your title.         %
% Make a linebreak in the title with         %
%                                            %
%            \newtitleline                   %
%                                            %
%%%%%%%%%%%%%%%%%%%%%%%%%%%%%%%%%%%%%%%%%%%%%%
%Comments on the use of buckow1.sty for
%\newtitleline
%writing your Buckow paper
%%%%%%%%%%%%%%%%%%%%%%%%%%%%%%%%%%%%%%%%%%%%%%
%}
\def\authors{Bernard de Wit}
\def\addresses{
Institute for Theoretical Physics,
Utrecht University\\
Princetonplein 5, 3508 TA Utrecht, The Netherlands }
\def\abstracttext{
We review recent developments in the theory of supermembranes and 
their relation to matrix models. \\
{\small\small\small
\noindent Invited talk given at the $31^{\rm st}$ 
International Symposium Ahrenshoop on the Theory of Elementary 
Particles, Buckow, September 2 - 6, 1997 (to appear in 
Fortschritte der Physik) and at the CERN Theory Division,  
December 3, 1997.}   
}
\makefront
%%%%%%%%%%%%%%%%%%%%%%%%%%%%%%%%%%%%%%%%%%%%%%%%
%                                              %
%  Insert now the remaining parts of           %
%  your article.                               %
%                                              %
%%%%%%%%%%%%%%%%%%%%%%%%%%%%%%%%%%%%%%%%%%%%%%%%

\section{Supersymmetric quantum mechanics}       
Consider the class of supersymmetric Hamiltonians of the form 
\begin{equation}\label{hamiltonian}
H = \frac{1}{g}{\rm Tr}\Big[ \ft{1}{2}{\bf P}^2 - \ft{1}{4}[X^a,X^b]^2
+\ft12 g\,\theta^{\rm T} \gamma_a [ X^a, \theta ]\,\Big] \, ,
\end{equation}
depending on a number of $d$-dimensional coordinates ${\bf 
X}=(X^1, \ldots,X^d)$, corresponding momenta ${\bf P}$, as well as real 
spinorial anticommuting coordinates $\theta_\alpha$, all taking values in 
the matrix representation of some Lie algebra. The phase space is 
restricted to the subspace invariant  
under the corresponding (compact) Lie group and is therefore subject 
to Gauss-type constraints. The above Hamiltonians arise in the 
zero-volume limit of supersymmetric Yang-Mills theories, which 
explains the presence of these constraints.         

The theories based on \eqn{hamiltonian} were proposed long ago as 
extended models of  
supersymmetric quantum mechanics with more than four 
supersymmetries \cite{CH}. The spatial dimension $d$ and the 
corresponding spinor dimension are severely restricted. The models exist 
for $d=2,3,5$, or 9 dimensions; the (real) spinor dimension equals 
$2, 4, 8$, or 16, respectively. Naturally this is also the number of 
independent supercharges. In what follows we restrict ourselves to 
the highest-dimensional case, where the model contains 16 
supercharges. However, additional charges can be obtained by  
splitting off an abelian factor of the gauge group (we will 
mainly consider the gauge group U$(N)$),  
\be
Q^+ = {\rm Tr}\Big[(P^a \gamma_a +\ft12 i [X^a,
X^b]\gamma_{ab})\,\theta\,\Big]\,,  \qquad
Q^- = {g}\; {\rm Tr} \left[\, \theta \,\right] \,.
\label{s-charges}
\ee
The $Q^+$ generate the familiar supersymmetry algebra (in the 
group-invariant subspace),
\be
\{Q^+_\alpha,Q^+_\beta\}\approx  H\, \delta_{\alpha\beta} \,
.\label{susyalg} 
\ee

A central theme of this lecture is that the supermembrane in the 
light-cone formulation is described by a quantum-mechanical model 
of the type above with an infinite-dimensional gauge group   
corresponding to the area-preserving diffeomorphisms of the 
membrane spacesheet \cite{DWHN}; the coupling constant $g$ is 
then equal to the total light-cone  
momentum $(P_-)_0$, which in a flat target space equals $P^+_0$. In 
11 spacetime dimensions the supermembrane  
is subject to 32 supercharges. The 16 charges $Q^-$ given in 
\eqn{s-charges}
are then associated with the center-of-mass superalgebra. The 
connection with the supermembrane shows that the manifest SO(9) 
symmetry, which from the  
viewpoint of the supermembrane is simply the exact transverse 
rotational invariance of the lightcone formulation, extends to 
the 11-dimensional Lorentz group in the limit of an appropriate  
infinite-dimensional gauge group \cite{DWMN,EMM}.  
 
Classical zero-energy configurations require all commutators
$[X^a,X^b]$ to vanish. Dividing 
out the gauge group implies that  
zero-energy configurations are thus 
parametrized by ${\bf R}^{9N}/S_N$. The zero-energy 
valleys in the potential extend all the way to infinity where  
they become increasingly narrow. Their existence raises questions 
about the nature of the spectrum of the Hamiltonian 
\eqn{hamiltonian}. In the bosonic versions of these models the 
wave function cannot freely extend to infinity, because at 
large distances it becomes more and more squeezed 
in the valley. By the uncertainty principle, this gives rise to 
kinetic-energy contributions which increase monotonically  
along the valley. Another way to see this effect 
is by noting that oscillations perpendicular to the valleys 
give rise to a zero-point energy, which induces an effective 
potential barrier that confines the wave function. 
This confinement causes the spectrum to be discrete. However, for 
the supersymmetric   
models defined by \eqn{hamiltonian} the situation is different. 
Supersymmetry can cause a cancelation of the transverse 
zero-point energy. Then the wave function is no longer confined, 
indicating that the supersymmetric models have a 
continuous spectrum. The latter was rigourously proven for the 
gauge group SU($N$) \cite{DWLN}.    
 
For the supermembrane, the classical zero-mass configurations 
correspond to zero-area stringlike configurations of arbitrary 
length. As the supermembrane mass is described by a 
Hamiltonian of the type \eqn{hamiltonian}, the mass spectrum of the 
supermembrane is continuous for the same reasons as given 
above. For a supermembrane moving in a target space with  
compact dimensions, winding may raise the mass of the membrane 
state. This is so because winding in more than one direction 
gives rise to a nonzero central charge in the  
supersymmetry algebra, which sets a lower limit on  
the membrane mass. 
This fact should not be interpreted as an indication that 
the spectrum becomes discrete. The possible 
continuity of the spectrum hinges on the two features mentioned 
above. First the system should possess  
continuous valleys of classically degenerate states. 
Qualitatively one recognizes immediately that this feature is not 
directly affected by winding. A classical membrane with 
winding can still have stringlike configurations of arbitrary length,  
without increasing its area. Hence the classical instability 
persists. The second feature is supersymmetry. 
Without winding it is clear that the valley configurations are 
supersymmetric, so that one concludes that the spectrum is 
continuous. With winding the latter aspect is more 
subtle. However, we note that, when the winding density is
concentrated in one part of the spacesheet, then valleys can
emerge elsewhere
corresponding to stringlike configurations with supersymmetry.
Hence, as a space-sheet local field theory,  
supersymmetry can be broken in one region where the winding is 
concentrated and unbroken in 
another. In the latter  region stringlike configurations can form, 
which, at least semiclassically, will not be suppressed by 
quantum corrections \cite{DWPP1}. 
However, in this case we can only describe the 
generic features of the spectrum. Our arguments do not  
preclude the existence of mass gaps. 

Finally, whether or not the Hamiltonian 
\eqn{hamiltonian} allows  normalizable or 
localizable zero-energy states, superimposed on the continuous  
spectrum, is a subtle question. Early discussion on the existence 
of such zero-energy states can be found in \cite{DWHN,DWN}; more 
recent discussions can be found in \cite{FH,mboundstates}. 
According to \cite{mboundstates} such states do indeed exist in 
$d=9$. There is an important difference  
between states whose energy is exactly equal to zero and states 
of positive energy. The supersymmetry algebra implies that 
zero-energy states are annihilated by the supercharges. 
Hence, they are supersinglets. The positive-energy states, on the 
other hand, must constitute full supermultiplets. So they are 
multiplets consisting of multiples of  $1+1$, $2+2$, $8+8$, or 
$128+128$ bosonic $+$ fermionic states, corresponding to $d=2,3,5$ 
or 9, respectively. 

To prove or disprove the existence of discrete states with 
winding is even more difficult. While the contribution of 
the bosonic part of the Hamiltonian increases by concentrating the 
winding density on part of the spacesheet, the matrix elements in 
the fermionic directions will also grow large, making it 
difficult to estimate the eigenvalues. At this moment the only 
rigorous result is the BPS bound that follows from the supersymmetry 
algebra. Obviously, the state of  lowest mass for 
given winding numbers is always a BPS state, which is invariant under some
residual supersymmetry. The counting of states proceeds in a way that is
rather similar to the case of no winding.

%%%%%%%%%%%%%%%%%%%%%%%%%%%%%%%%%%%%%%%%%%%%%%%%%%%%%%%%%%%%%%%%
\section{Supermembranes}                         
Fundamental supermembranes can be described in terms of actions 
of the Green-Schwarz type, possibly in a nontrivial but 
restricted (super)spacetime background \cite{BST}. Such actions exist 
for supersymmetric $p$-branes, where $p= 0$, 1, $\ldots$ defines the
spatial dimension of the brane. Thus for
$p=0$ we have a superparticle, for $p=1$ a superstring, for
$p=2$ a supermembrane, and so on. The
dimension of spacetime in which the superbrane can live is
very restricted. These restrictions arise from the 
fact that the action contains a 
Wess-Zumino-Witten term, whose supersymmetry depends sensitively
on the spacetime dimension. If the coefficient of this term
takes a particular value then the action possesses an additional
fermionic gauge symmetry, the so-called $\kappa$-symmetry.
This symmetry is necessary to
ensure the matching of (physical) bosonic and fermionic
degrees of freedom.
In the following we restrict ourselves to 
supermembranes (i.e., $p=2$) in 11 dimensions.  

The supermembrane action \cite{BST} is written in terms of 
superspace embedding coordinates $Z^M(\zeta)=(X^\mu(\zeta),
\theta(\zeta))$, which are functions of the three 
world-volume coordinates $\zeta^i$ ($i = 0$, $1$, $2$). It takes 
the following form, 
\be
S[Z(\zeta)] =\int {\rm d}^3\zeta \;\Big[- \sqrt{-g(Z(\zeta))} - \ft13
\varepsilon^{ijk}\, \Pi_i^A \Pi_j^B\Pi_k^C \,B_{CBA}(Z(\zeta)) 
\,\Big]\,,
\label{supermem}
\ee
where $\Pi^A_i = \pa Z^M/\pa\zeta^i \; E_M^{\;A}$ and the 
induced metric equals $g_{ij} =\Pi^r_i \Pi^s_j \,\eta_{rs}$, with
$\eta_{rs}$ the constant Lorentz-invariant metric. 
Flat superspace is characterized by
\bea\label{flatssquantities}
E_\mu{}^r &\!\!=\!\!& \d_\mu{}^r \, , \hspace{43mm} E_\mu{}^a =0  \, ,\nonumber \\
E_\alpha{}^a &\!\!=\!\!& \delta_\alpha{}^a  \, ,\hspace{42.6mm} 
E_\alpha{}^r = -(\bar\theta \Gamma^r)_{\alpha} \, ,\nonumber \\
B_{\mu\n\alpha} &\!\!=\!\!& (\bar\theta\Gamma_{\m\n})_{\alpha}\,
, \hspace{32.65mm}
B_{\m\a\b} = 
(\bar\theta\Gamma_{\m\n})_{(\a}\,(\bar\theta\Gamma^\n)_{\b)} \,,\nonumber \\
 B_{\a\b\g} &\!\!=\!\!& (\bar\theta\Gamma_{\m\n})_{(\a}\, 
(\bar\theta\Gamma^\m)_{\b}\, 
(\bar\theta\Gamma^\n)_{\g)}\,,\hspace{6.2mm}  
B_{\mu\nu\rho} =0  \, .  
\eea
The gamma matrices 
are denoted by $\Gamma^r$; gamma matrices 
with more than one index denote antisymmetrized products of gamma
matrices with unit weight. In flat  
superspace the supermembrane Lagrangian, written in 
components, reads (in the notation 
and conventions of \cite{DWHN}),  
\bea
&&{\cal L} =- \,\sqrt{-g(X,\theta)} - \varepsilon^{ijk} \,
\overline\theta\Gamma_{\mu\nu}\partial_k\theta \Big [{\textstyle{1\over 2}}
\,\partial_i X^\mu (\partial_j X^\nu +
\overline\theta\Gamma^\nu\partial_j \theta) + {\textstyle{1\over 6}}
\,\overline\theta\Gamma^\mu\partial_i\theta\;
\overline\theta\Gamma^\nu\partial_j\theta \Big] \,, \qquad{}
\label{action}
\eea
The target space can 
have compact dimensions which permit winding membrane states 
\cite{DWPP1}. In flat superspace the induced metric,
\be
g_{ij} = (\partial_iX^\mu + \overline\theta\Gamma^\mu
\partial_i\theta) (\partial_jX^\nu +
\overline\theta\Gamma^\nu
\partial_j\theta) \,\eta_{\mu\nu}\,,
\ee
is supersymmetric. Therefore the first term in 
\eqn{action} is trivially invariant under spacetime
supersymmetry. In $4$, $5$, $7$, or $11$ spacetime dimensions the 
second term proportional to $\varepsilon^{ijk}$ is also 
supersymmetric (up to a total divergence) and the full action is 
invariant under  $\kappa$-symmetry.
 
In the case of the open supermembrane, $\kappa$-symmetry imposes 
boundary conditions on the fields \cite{open}. They must 
ensure that the following integral over the  
boundary of the membrane world volume vanishes,
\bea
&&\int_{\del M} \Big[ \ft12{\rm d} X^\m \wedge ( {\rm d}X^\n + 
\bar\theta\G^\nu {\rm d}\theta)\, \bar \theta \G_{\m\n} \d_\kappa 
\theta  +\ft16\bar\theta\G^\mu {\rm d}\theta\wedge \bar\theta\G^\nu {\rm 
d}\theta\, \bar\theta\G_{\m\nu} \d_{\kappa}\theta \nonumber\\
&&\hspace{40mm} +\ft12( {\rm d}X^\m - \ft13 \bar\theta\G^\mu {\rm 
d}\theta ) \wedge 
\bar\theta\G_{\m\nu} {\rm d}\theta\; \bar\theta\G^\n \d_{\kappa}\theta
\Big ]=0 \,.
\eea
This can be achieved by having a ``membrane D-$p$-brane'' at 
the boundary with $p=1,5$, or 9, which is defined in terms of 
$(p+1)$ Neumann and $(10-p)$ Dirichlet boundary 
conditions for the $X^\mu$, together with corresponding boundary 
conditions on the fermionic coordinates.  
More explicitly, we define projection operators
\be 
{\cal P}_\pm=\ft12\Big({\bf 1} \pm \G^{p+1}\, \G^{p+2}\cdots 
\G^{10}\Big)\,, \label{projectors}
\ee
and impose the Dirichlet boundary conditions
\bea
\del_\parallel \, X^M\big|&=& 0\,, \qquad M=p+1,\ldots,10\,, \nonumber\\
{\cal P}_- \q\big|&=&0\, , \label{boundcond}
\eea 
where $\del_\perp$ and $\del_\parallel$ define the world-volume 
derivatives perpendicular or tangential to the surface swept out 
by the membrane boundary in the target space. Note that the fermionic 
boundary condition implies that ${\cal P}_- \del_\parallel\q=0$. 
Furthermore, it implies that spacetime 
supersymmetry is reduced to only 16 supercharges associated with 
spinor parameters ${\cal P}_+\epsilon$, which is {\it chiral} with 
respect to the ($p+1$)-dimensional world volume of the 
D-$p$-brane at the boundary. With respect to 
this reduced supersymmetry, the superspace coordinates decompose 
into two parts, one corresponding to $(X^M, {\cal P}_-\theta)$ and the 
other corresponding to $(X^m, {\cal P}_+\theta)$ where
$m=0,1,\ldots,p$. While for the 
five-brane these superspaces exhibit a somewhat balanced decomposition in 
terms of an equal number of bosonic and fermionic coordinates,
the situation for $p=1,9$ shows heterotic features in that 
one space has an excess of fermionic and the other an excess of 
bosonic coordinates. Moreover, we note that supersymmetry may be further
broken, e.g.\ by choosing different Dirichlet conditions 
on nonconnected segments of the supermembrane boundary.

The Dirichlet boundary conditions can be supplemented by the 
following Neumann boundary conditions,
\bea
\del_\perp \, X^m\big|&=& 0 \qquad m=0,1,\ldots,p \,,\nonumber\\
{\cal P}_+ \del_\perp \q \big|&=&0 \,. \label{Nboundcond}
\eea 
These do not lead to a further breakdown of the rigid spacetime 
symmetries.

We now continue and follow the light-cone quantization described 
in \cite{DWHN}. The supermembrane Hamiltonian takes the form 
\begin{eqnarray}
\label{memham}
H&=&  \frac{1}{P_0^+}\, \int {\rm d}^2\s \, \sqrt{w}\,
\bigg[ \, \frac{P^a\, P_a }{2\,w} + \ft{1}{4} \{\, 
X^a,X^b\,\}^2 -P^+_0\, \qb\,\g_- 
\g_a\, \{\, X^a , \q\,\}\, \bigg]\, .
\end{eqnarray}
Here the integral runs over the spatial components of the
world volume denoted by $\s^1$ and $\s^2$, while 
$P^a(\s)$ ($a=2,\ldots,9$) are the momenta conjugate to the 
transverse coordinates $X^a$. In this gauge 
the light-cone coordinate $X^+=(X^{1}+X^0)/\sqrt2$ is linearly related to the 
world-volume time denoted by $\tau$. The momentum $P_-$ is time 
independent and proportional to the center-of-mass value $P^+_0= 
(P_-)_0$ times some  
density ${\sqrt{w(\s)}}$ of the spacesheet, whose spacesheet 
integral is normalized to unity. The center-of-mass momentum
$P_0^-$ is equal to minus the Hamiltonian \refer{memham}
subject to the gauge 
condition \tmath{\g_+\, \q=0}. And finally we made use of the
Poisson bracket \tmath{\{ A,B\} } defined by
\be
\{ A(\s ),B(\s )\} = \frac{1}{\sqrt{w(\s)}}\, \varepsilon^{rs}\, 
\del_r A(\s )\, 
\del_s B(\s ). \label{poisbrak}
\ee
Note that the coordinate $X^-=(X^{1}-X^0)/\sqrt2$ itself does
not appear in the Hamiltonian \refer{memham}. It is defined via 
\be
P^+_0\, \del_rX^-= - \frac{{\bf P} \cdot \del_r{\bf X}}{\sqrt{w}} - 
P^+_0\, \qb\g_-\del_r\q\,, \label{delxminus}
\ee
and implies a number of constraints that will be important in 
the following. Obviously, the right-hand side of \eqn{delxminus} 
must be closed; without winding in $X^-$, it must be 
exact. 

The equivalence of the large-$N$ limit of SU$(N)$ quantum mechanics
with the closed supermembrane model is based on the residual invariance
of the supermembrane action in the light-cone gauge. This 
invariance corresponds to the area-preserving diffeomorphisms of 
the membrane surface. 
These are defined by transformations of the worldsheet coordinates
\begin{equation}
\s^r \to \s^r + \xi^r(\s) \,, 
\end{equation}
with
\begin{equation}\label{APD}
\del_r(\sqrt{w(\s)}\, \xi^r(\s)\, )=0.
\end{equation}
It is convenient to rewrite this condition in terms of dual spacesheet 
vectors by  
\be                                  
\sqrt{w(\s)}\,\xi^r(\s)= \varepsilon^{rs}\, F_s(\s)\, .\label{1form}
\ee
In the language of differential forms the
condition \refer{APD} may then be simply recast as \tmath{{\rm 
d}F=0}. The trivial solutions are the exact forms \tmath{F={\rm 
d}\xi}, 
or in components,  
\be
F_s=\del_s\xi(\s)\,,\label{exact}
\ee
for any globally defined function $\xi(\s)$. The nontrivial solutions are
the closed forms which are not exact. On a Riemann surface of
genus $g$ there are precisely $2g$ linearly independent non-exact 
closed forms, whose integrals along the homology cycles are 
normalized to unity\footnote{%
  In the mathematical literature the globally defined exact forms 
  are called ``hamiltonian vector fields'', whereas the closed 
  but not exact forms which are not globally defined go under the 
  name ``locally hamiltonian vector fields''.}. %
In components we write
\be
F_s=\phi_{(\l)\, s}\;, \qquad \l=1,\ldots,2g\,.
\ee

The commutator of two infinitesimal area-preserving 
diffeomorphisms is determined by the product rule
\begin{equation}
\xi_r^{(3)} = \partial_r \left( \frac{\epsilon^{st}}{\sqrt{w}} 
\xi_s^{(2)}\xi_t^{(1)}\right) \,,
\end{equation}
where both $\xi_r^{(1,2)}$ are closed vectors. Because 
$\xi_r^{(3)}$ is exact, the exact vectors thus 
generate an invariant subgroup of the area-preserving 
diffeomorphisms. As we shall discuss in the next section this 
subgroup can be approximated by SU$(N)$ in the large-$N$ limit, 
at least for closed membranes. For open membranes the boundary conditions 
on the fields \refer{boundcond} lead to a smaller group, such as SO($N$).

The presence of the closed but non-exact forms is crucial for 
the winding of the embedding coordinates. More precisely, while 
the momenta ${\bf P}(\s)$ and the fermionic coordinates 
$\theta(\s)$ remain single valued on the spacesheet, the 
embedding coordinates, written as one-forms with components 
$\del_r {\bf X}(\s)$ and  $\del_r X^-(\s)$, are decomposed into 
closed one-forms. Their non-exact contributions are multiplied by an 
integer times the length of the compact direction.
The constraint alluded to above 
amounts to the condition that the right-hand side of 
\refer{delxminus} is closed. 

Under the full group of area-preserving diffeomorphisms the fields $X^a$,
$X^-$ and $\q$ transform according to
\be
\label{APDtrafoXtheta}
\d X^a= \displaystyle{\varepsilon^{rs}\over \sqrt{w}}\, \xi_r\, \del_s X^a\,,
\quad 
\d X^-= \displaystyle {\varepsilon^{rs}\over \sqrt{w}}\, \xi_r\, \del_s 
X^-\,, 
\quad
\d \q^a= \displaystyle {\varepsilon^{rs}\over \sqrt{w}}\, 
\xi_r\, \del_s \q\,, 
\ee
where the time-dependent reparametrization $\xi_r$ consists of
closed exact and non-exact parts. Accordingly there is a gauge
field $\o_r$, which is therefore closed as well and transforming 
as 
\be\label{APDtrafoomega}
\d\o_r=\del_0\xi_r + \del_r \bigg( {\varepsilon^{st}\over\sqrt{w}}\,
\xi_s\,\o_t\bigg)\,.
\ee
Corresponding covariant derivatives are 
\be
\label{covderiv}
D_0 X^a = \displaystyle\del_0X^a - {\varepsilon^{rs}\over \sqrt{w}}\, 
\o_r\, \del_s X^a\,, \qquad  
D_0 \q  = \displaystyle \del_0\q - {\varepsilon^{rs}\over \sqrt{w}}\, 
\o_r\, \pa_s\q\,, 
\ee 
and likewise for \tmath{D_0 X^-}. 

The action corresponding to the following Lagrangian density is
then gauge invariant under the 
transformations \refer{APDtrafoXtheta} and \refer{APDtrafoomega},
\bea
\label{gtlagrangian}
{\cal L}&=&P^+_0\,\sqrt{w}\, \Big[\,  
\ft{1}{2}\,(D_0{\bf X})^2 + \qb\,\g_-\,
D_0\q - \ft{1}{4}\,(P^+_0)^{-2}\,  \{ X^a,X^b\}^2 \\
&& \hspace{16mm}  + (P^+_0)^{-1}\, \qb\,\g_-\,\g_a\,\{X^a,\q\} +
  D_0 X^-\Big]\, ,\nonumber 
\eea
where we draw attention to the last term proportional to
$X^-$, which can be dropped in the absence of winding. 
Moreover, we note that for open
supermembranes, \refer{gtlagrangian} is invariant under the 
transformations \refer{APDtrafoXtheta} and \refer{APDtrafoomega} 
only if $\xi_\parallel=0$ holds on the boundary.
This condition defines a subgroup of the 
group of area-preserving transformations, which is consistent 
with the Dirichlet conditions \refer{boundcond}. Observe that 
here $\del_\parallel$ and $\del_\perp$ refer to the {\it 
spacesheet} derivatives tangential and perpendicular to the 
membrane boundary\footnote{%
  Consistency of the Neumann boundary conditions 
  \refer{Nboundcond} with the area-preserving diffeomorphisms 
  \refer{APDtrafoXtheta} further imposes 
  $\partial_\perp\xi^\parallel=0$ 
  on the boundary, where indices are raised according to 
  \refer{1form}.}. % 

The action corresponding to
\refer{gtlagrangian} is also invariant under the 
supersymmetry transformations  
\bea
\d X^a &=& -2\, \bar{\e}\, \g^a\, \q\,, \nonumber\\
\d \q  &=& \ft{1}{2} \g_+\, (D_0 X^a\, \g_a + \g_- )\, \e 
+\ft{1}{4}(P^+_0)^{-1} \, \{ X^a,X^b \}\, \g_+\, \g_{ab}\, \e ,
\nonumber\\ 
\d \o_r &=& -2\,(P^+_0)^{-1}\, \bar{\e}\,\pa_r\q\,.
\label{susytrafos}
\eea
The supersymmetry variation of $X^-$ is not relevant and may be
set to zero. For the open case one finds that the boundary conditions
$\omega_\parallel=0$
and \mbox{$\epsilon={\cal P}_+\,\epsilon$} must be fulfilled
in order for \refer{susytrafos} to be a symmetry of the action.
In that case the theory takes the form of a gauge theory coupled 
to matter. The pure gauge theory is associated with the Dirichlet 
and the matter with the Neumann (bosonic and 
fermionic) coordinates.  

In the case of a `membrane D-$9$-brane' one now sees that the
degrees of freedom  on the `end-of-the world' $9$-brane precisely
match those of 10-dimensional heterotic strings. {\it On} the boundary
we are left with eight propagating bosons $X^m$ (with $m=2,
\ldots,9$), as $X^{10}$ is constant on the boundary 
due to \refer{boundcond},
paired with the 8-dimensional chiral spinors $\theta$ (subject 
to $\g_+ \theta= {\cal P}_-\theta=0$),
i.e., the scenario of Ho\u{r}ava-Witten \cite{horvw}.

The full equivalence with the membrane Hamiltonian is now established by
choosing the $\o_r=0$ gauge and passing to the Hamiltonian 
formalism. The field equations for $\o_r$ then lead to
the membrane constraint \refer{delxminus} (up to exact contributions), 
partially defining \tmath{X^-}.
Moreover the Hamiltonian corresponding to the gauge theory Lagrangian of 
\refer{gtlagrangian} is nothing
but the light-cone supermembrane Hamiltonian \refer{memham}.
Observe that in the above gauge theoretical construction the space-sheet
metric $w_{rs}$ enters only through its density $\sqrt{w}$ and hence
vanishing or singular metric components do not pose problems.

We are now in a position to study the full 11-dimensional supersymmetry
algebra of the winding supermembrane. For this we decompose the
supersymmetry charge $Q$ associated with the transformations 
\refer{susytrafos}, into two 16-component spinors,
\be
Q= Q^+ + Q^- , \quad \mbox{where}\quad
 Q^\pm = \ft{1}{2}\, \g_\pm\,\g_\mp\, Q\,, \label{Qdecomposition}
\ee
to obtain
\bea
Q^+&=&\int {\rm d}^2 \s \, \Big(\, 2\, P^a\, \g_a + \sqrt{w}\, \{\,
X^a, X^b\, \} \, \g_{ab}\, \Big) \, \q \,, \nonumber \\
Q^-&=& 2\, P^+_0\, \int {\rm d}^2\s\, \sqrt{w}\, \g_-\, \q .
\label{Q-cont}
\eea
In the presence of winding the supersymmetry algebra takes the 
form \cite{DWPP1}
\begin{eqnarray}
\label{contsusy}
(\, Q^+_\a, \bar{Q}^+_\b\, )_{\mbox{\tiny DB}} &=& 2\, 
(\g_+)_{\a\b}\, H 
 - 2\,  (\g_a\, \g_+)_{\a\b}\, \int{\rm d}^2\s\, \sqrt{w}\, \{\, 
X^a, X^-\,\}\, , \nonumber \\ 
(\, Q^+_\a, \bar{Q}^-_\b\, )_{\mbox{\tiny DB}} &=& -(\g_a\,\g_+\,
\g_- )_{\a\b}\, P^a_0  
  - \ft{1}{2}\,(\g_{ab}\, \g_+\g_- )_{\a\b}\, \int {\rm d}^2\s\,
\sqrt{w}\, \{\, X^a,X^b\,\}\,,\nonumber\\[1mm] 
(\, Q^-_\a, \bar{Q}^-_\b\, )_{\mbox{\tiny DB}} &=& -2\, (\g_- )_{\a\b}\, P^+_0\, , 
\end{eqnarray}
where use has been made of the Dirac brackets of the phase-space 
variables and the defining equation \refer{delxminus} for $\pa_r X^-$.

The new feature of this supersymmetry algebra is the emergence of the 
central charges in the first two anticommutators, which are
generated through the winding contributions.
They represent topological quantities obtained by integrating
the winding densities 
\begin{equation}
z^{a}(\s)=\varepsilon^{rs}\,\del_r X^a\,\del_s X^-
\end{equation}
and
\begin{equation}
z^{ab}(\s) =\varepsilon^{rs}\,\del_r X^a\,\del_s X^b
\end{equation}
over the space-sheet. It is gratifying to observe the manifest
Lorentz invariance of \refer{contsusy}. Here  we should point out
that, in adopting the light-cone gauge, we assumed that there was
no winding for 
the coordinate $X^+$. In \cite{BSS} the corresponding algebra for
the matrix regularization was studied. 
The result coincides with ours in the
large-$N$ limit, in which an additional longitudinal five-brane
charge vanishes, provided that one identifies the longitudinal
two-brane charge with the central charge in the
first line of \refer{contsusy}. This identification requires the definition of
$X^-$ in the matrix regularization, a topic that we return to in 
the next section. The form of the algebra is another indication of the 
consistency of the supermembrane-supergravity system.

Until now we discussed the general case of a flat target space 
with possible winding states. To make the identification with the 
matrix models more explicit, let is ignore the winding and 
split off the center-of-mass (CM) variables. First of all, the 
constant $P^+_0$  
represents the  membrane CM momentum in the 
direction associated with the coordinate $X^-$,
\be
P^+_0 = \int {\rm d}^2\! \sigma \,P^+ . 
\ee
The other CM coordinates and momenta are    
\be
{\bf P}_0 = \int {\rm d}^2\! \sigma \; {\bf P} \,,  \qquad {\bf 
X}_0 = \int \!{\rm d}^2\!\sigma\sqrt{w(\sigma)}\, {\bf X}(\sigma)\,, 
\qquad 
\theta_0 = \int \!{\rm d}^2\!\sigma 
\sqrt{w(\sigma)}\, \theta(\sigma)\,.  
\ee
In the light-cone gauge we are left with the transverse
coordinates $\bf X$ and corresponding momenta $\bf P$, which
transform as vectors under the SO(9) group of transverse
rotations. Only
sixteen fermionic components $\theta$ remain, which transform as
SO(9) spinors. Furthermore we have the CM momentum
$P_0^+$ and the CM coordinate $X^-_0$ (the
remaining modes in $X^-$ are dependent), while the CM momentum
$P_0^-$ is equal to minus the supermembrane Hamiltonian and takes 
the following form 
\be
H = {{\bf P}_0^{\,2}\over 2 P_0^+} + {{\cal M}^2\over 2P_0^+}  
\,. \label{mbhamiltonian}
\ee 
Here $\cal M$ is the supermembrane mass operator, which does {\it
not} depend on any of the CM coordinates or momenta. The explicit 
expression for ${\cal M}^2$ is 
\be
{\cal M}^2 =  \int {\rm d}^2\!\s \; \sqrt{w(\s)} \bigg[ {[{\bf 
P}^2(\s)]' \over w(\s)}  +\ft12 \Big(\{X^a,X^b\}\Big)^2 - 2 P_0^+ 
\, \bar\theta\g_- \g_a\{X^a,\theta\}\bigg]\,, \label{mass}
\ee
where $[{\bf P}^2]^\prime$ indicates that the contribution of the CM 
momentum ${\bf P}_0$ is suppressed. 

The structure of the Hamiltonian \eqn{mbhamiltonian} shows that the wave
functions for the supermembrane now factorize into a
wave function pertaining to the CM modes and a wave
function of the supersymmetric quantum-mechanical system that
describes the other modes. For the latter the mass operator plays 
the role of the Hamiltonian. When the mass operator vanishes on the state, 
then the 32 supercharges act exclusively on the CM coordinates 
and generate a massless supermultiplet of eleven-dimensional 
supersymmetry. In case there is no other degeneracy beyond that 
caused by supersymmetry, the resulting supermultiplet is the one 
of supergravity, describing the graviton, the antisymmetric tensor 
and the gravitino. In terms of the $SO(9)$ helicity 
representations, it  
consists of ${\bf 44} \oplus {\bf 84}$ bosonic and $\bf 128$
fermionic states. For an explicit construction of these states, 
see \cite{PW}. 
When the mass operator does not vanish on the states, we are 
dealing with huge supermultiplets consisting of multiples of 
$2^{15}+2^{15}$ states.

%%%%%%%%%%%%%%%%%%%%%%%%%%%%%%%%%%%%%%%%%%%%%%%%%%%%%%%%%%%%%%%%%%%%%
\section{The matrix approximation}

The expressions for the Hamiltonian \eqn{memham}, the 
supercharges \eqn{Q-cont} and the constraints associated with  
\eqn{delxminus} are clearly in direct 
correspondence with the Hamiltonian, supersymmetry charges and 
the Gauss constraints for the matrix models introduced in 
section~1. This correspondence between de supermembrane and 
supersymmetric quantum mechanics becomes exact after  
one replaces $P^+_0$ by the coupling constant $g$ and rewrites 
the spinor coordinates in terms of a real SO(9) spinor basis. 
In order to make the relation more explicit one may expand functions 
on the spacesheet in a complete set of functions $Y_A$ with 
$A= 0,1,2, \ldots, \infty$. It is convenient to choose $Y_0=1$. 
Furthermore we choose a basis of the closed one-forms, consisting 
of the exact ones, $\pa_r Y_A$, and a set of closed nonexact 
forms denoted by $\phi_{(\l)r}$. 
Completeness implies the following decompositions,
\bea
\{ Y_A, Y_B\} &=& f_{AB}{}^{\!C}\, Y_C\,, \nonumber  \\[1.9mm]
{\varepsilon^{rs}\over \sqrt w} \,\phi_{(\l)r}\,\pa_s Y_A &=& 
f_{\l A}{}^{\!B} \, Y_B\,, \nonumber\\
{\varepsilon^{rs}\over \sqrt w} \,\phi_{(\l)r}\,
\phi_{(\l^\prime)s} &=& f_{\l \l^\prime} {}^{\!A} \, Y_A\,,
\eea
so that the constants $f^{AB}{}_{\!C}$, $f_{\l A}{}^{\!B}$ and 
$f_{\l \l^\prime} {}^{\!A}$ represent the
structure constants of the infinite-dimen\-sional group of 
area-preserving diffeomorphisms. 
Lowering of indices can be done with the help of the 
invariant metric
\begin{equation}
\eta_{AB}= \int {\rm d}^2\s\, \sqrt{w(\s)}\; Y_A(\s)\, Y_B(\s)\,.
\end{equation}
There is no need to introduce a metric for the $\l$ indices. 
Observe that we have $\eta_{00}=1$. Furthermore it is convenient 
to choose the functions $Y_A$ with $A\geq 1$ such that 
$\eta_{0A}=0$. Completeness implies
\be
\eta^{AB}\,Y_A(\s)\,Y_B(\rho) = {1\over \sqrt{w(\s)}}\,
\d^{(2)}(\s,\rho)\,.
\ee

After lowering of upper indices, the structure constants are defined as 
follows \cite{DWMN,DWPP1}, 
\bea
f_{ABC} &=& \int {\rm d}^2\s\, \varepsilon^{rs}\,\del_r Y_A(\s)\, 
\del_s Y_B(\s)\, Y_C(\s)\,, \nonumber \\ 
f_{\l BC} &=& \int {\rm d}^2\s\, \varepsilon^{rs}\,\phi_{(\l)\, 
r}(\s)\, \del_s Y_B(\s)\, Y_C(\s)\,, \nonumber \\ 
f_{\l \l^\prime C} &=& \int {\rm d}^2\s\, \varepsilon^{rs}\,
\phi_{(\l)\, r}(\s)\, \phi_{(\l^\prime)\, s}(\s)\, Y_C(\s) \, . 
\eea 
Note that we have $f_{AB0}=f_{\l B0}=0$.

Using the above basis one may write down the following
mode expansions for the phase-space variables of the
supermembrane, 
\bea
\del_r{\bf X}(\s) &=& \sum_{\l} \,{\bf X}^\l\, \phi_{(\l)\, r}(\s) 
+ \sum_A\, {\bf X}^A\, \del_r Y_A(\s)\,,\nonumber \\
{\bf P}(\s) &=& \sum_A \,\sqrt{w(\s)}\; {\bf P}^A\, Y_A(\s)\,, \nonumber\\
\q(\s) &=& \sum_A \,\q^A\, Y_A(\s) \,, \label{modeexp}
\eea
introducing winding modes for the transverse coordinates $\bf X$. 
A similar expansion exists for $X^-$.  

Other tensors are needed, for instance, to write down the Lorentz 
algebra generators \cite{DWMN}. An obvious tensor is 
given by 
\be
d_{ABC} =  \int {\rm d}^2\s\, \sqrt{w(\s)}\; Y_A(\s)\, Y_B(\s)\, 
Y_C(\s) \,, 
\ee
which is symmetric in all three indices and satisfies $d_{AB0}= 
\eta_{AB}$. 
Another tensor, whose definition is more subtle, arises when 
expressing $X^-$ in terms of the other coordinates and momenta. 
We recall that $X^-$ is restricted by \eqn{delxminus}, which 
implies the following Gauss-type constraint, 
\be
\varphi^A= f_{BC}{}^{\!A}\Big[ {\bf P}^B\cdot {\bf X}^C + P_0^+\, \bar 
\theta^B\g_- \theta^C\Big] + f_{B\l}{}^{\!A} \, {\bf P}^B\cdot {\bf 
X}^\l \approx 0\,.  \label{constraint}
\ee
The coordinate $X^-$ receives contributions proportional to $Y_A(\s)$, 
which can be parametrized by ($A\not=0$)
\be
X^-_A\approx  {1\over 2P_0^+}\, c^A{}_{\!BC}\Big[{\bf P}^B\cdot{\bf X}^C + 
P_0^+\, \bar  \theta^B\g_- \theta^C\Big]  + {1\over 2P_0^+}\, 
c^A{}_{\!B\l}\, {\bf P}^B\cdot{\bf X}^\l\,. \label{X-A}
\ee
In addition $X^-$ has CM and winding modes. Observe that the 
tensors $c^A{}_{\!BC}$ and $c^A{}_{\!B\l}$
are somewhat ambiguous, as \eqn{X-A} is only defined up to the 
constraints \eqn{constraint}. The symmetric component of 
$c^A{}_{\!BC}$ is, however, fixed and given by  
$c^A{}_{\!BC}+c^A{}_{\!CB}= 
-2 d_{ABC}$. Note that $c^A{}_{\!B0}=0$. There are many other 
identities between the various tensors, such as \cite{DWMN}, 
\bea
&&f_{[AB}{}^{\!E}\, f_{C]E}{}^{\!D} =d_{(AB}{}^{\!E}\, 
f_{C)E}{}^{\!D} = d_{ABC} 
\, f_{[DE}{}^{\!B}\, f_{FG]}{}^{\!C}= \nonumber\\
&&c_{DE}{}^{\![A}f^{BC]E}= 
d_{EA[B}d_{C]D}{}^{\!E} = 0\,. \label{identities}
\eea

If we replace the group of the area-preserving diffeomorphisms   
by a finite group, then \eqn{mass} defines the 
Hamiltonian of a supersymmetric quantum-mechanical system based 
on a finite number of degrees of freedom \cite{GoldstoneHoppe}.
In the limit to the infinite-dimensional group we thus recover the
supermembrane. This observation enables one to
regularize the supermembrane in a
supersymmetric way by considering a limiting procedure based on a
sequence of groups whose limit yields the area-preserving 
diffeomorphisms. For closed  
membranes of certain topology it is known how to approximate a 
(sub)group of the area-preserving diffeomorphisms as a particular $N\to 
\infty$ limit of SU($N$). To  
be precise, it can be shown that the structure constants of 
SU($N$) tend to those of the diffeomorphism subgroup associated with 
the hamiltonian vectors, up to corrections of order  
$1/N^2$. While some of the identities \eqn{identities} remain valid 
at finite $N$, others receive corrections of order 
$1/N^2$. Furthermore, the tensors 
$c^A{}_{\!BC}$ and $c^A{}_{\!B\l}$ are intrinsically undefined at 
finite $N$. Therefore, the expression for $X^-$ is ambiguous for 
the matrix model and Lorentz invariance holds only 
in the large-$N$ limit \cite{DWMN,EMM}. 

The nature of the large-$N$ limit itself is subtle and depends 
on the membrane topology. As long as $N$ is finite, no 
distinction can be made with regard to the topology. In some 
sense, all topologies are thus included at the level of finite 
$N$. However, the diffeomorphisms associated with the harmonic 
vectors are problematic, because they cannot be incorporated for 
finite $N$, at least not at the level of the Lie algebra. This 
was shown in \cite{DWMN}, where it was established that the 
finite-$N$ approximation to the structure constants \tmath{f_{\l 
BC}}  violates the Jacobi identities for a toroidal membrane. 
Therefore it seems impossible to present a matrix
model regularization of the supermembrane with winding 
contributions. There exists a standard prescription for dealing 
with matrix models with winding \cite{Tdual}, however, which 
is therefore conceptually different. The  
consequences of this difference are not well understood. The 
prescription amounts to adopting the gauge group $[{\rm 
U}(N)]^M$, for winding in one dimension, which in the limit $M\to 
\infty$ leads to supersymmetric Yang-Mills
theories in $1+1$ dimensions \cite{Tdual}. Hence, in this way it 
is possible to extract extra dimensions from a suitably chosen 
infinite-dimensional gauge group. Obviously this approach can be 
generalized to a hypertorus.   

Finally we add that the matrix regularization works also for the 
case of open supermembranes. In that case one deals with certain 
subgroups of SU($N$). We refer to \cite{open} for 
further details.

%%%%%%%%%%%%%%%%%%%%%%%%%%%%%%%%%%%%%%%%%%%%%%%%%%%%%%%%%%
\section{Membranes and matrix models in curved space}

So far we considered a supermembrane moving in a flat 
target superspace. To that order we substituted the flat superspace 
expressions \eqn{flatssquantities} into the supermembrane 
action \eqn{supermem}. However, these expression can in 
principle be evaluated for nontrivial backgrounds, such as
those induced by a nontrivial target-space metric, a target-space 
tensor field and a target-space gravitino field, corresponding to 
the fields of (on-shell) 11-dimensional supergravity. This
background can in principle be incorporated into superspace by a
procedure known as `gauge completion'~\cite{gc}. For 11-dimensional 
supergravity, the first steps of this procedure have been carried 
out long ago~\cite{CF}, but unfortunately only to first order in fermionic 
coordinates $\theta$. 

For brevity of the presentation, let us 
just confine ourselves to the purely bosonic case and present
the light-cone formulation of the membrane in a 
background consisting of the metric $G_{\mu\nu}$ and the tensor 
gauge field $C_{\mu\nu\rho}$ \cite{DWPP3}.
The Lagrangian density for the bosonic membrane follows directly 
{}from \eqn{supermem},  
\begin{equation}
{\cal L} = -\sqrt{-g} + \ft{1}{6}\varepsilon^{ijk} \partial_i X^\mu\,
\partial_j
X^\nu \,\partial_k X^\rho\, C_{\mu\nu\rho} \, ,
\end{equation}
where $g_{ij}= \pa_i X^\m \,\pa_j X^\n \,\eta_{\m\n}$.  
For the light-cone formulation, the coordinates are treated in 
the usual fashion in terms of light-cone coordinates $X^\pm$ and 
transverse coordinates $\bf X$. Furthermore we use the 
diffeomorphisms in the target space to bring the metric in a convenient form~\cite{GS},
\begin{equation}
\label{metricgauge}
G_{--}=G_{a-}=0\, .
\end{equation}
Following the same steps as for the membrane in flat 
space, discussed in section~2, one again derives a Hamiltonian 
formulation. Interestingly enough, the constraint 
takes the same form as \eqn{delxminus}. 
Of course, the definition of the momenta in terms of the 
coordinates and their derivatives does involve the 
background fields, but at the end all explicit dependence 
on the background cancels out. 

The Hamiltonian now follows straightforwardly. After additional 
gauge choices, 
\begin{equation}
C_{+-a} = 0\, , \quad C_{-ab}=0\, ,\quad G_{+-}=1\, ,
\end{equation}
it takes the form
\bea
H&=& \int {\rm d}^2\sigma\, \Big \{
\frac{1}{P_-}\Big[\ft{1}{2}(P_a-C_a-P_-\, G_{a+})^2+ \ft{1}{4}
(\varepsilon^{rs}\, \partial_r X^a\, \partial_s X^b )^2\Big]\nonumber\\
&&\hspace{13.5mm}  -\ft{1}{2} P_-\, G_{++}-  \ft{1}{2}\varepsilon^{rs}\, 
\partial_r X^a\, \partial_s X^b\, C_{+ab} \Big \}\,,
\eea
We want to avoid explicit time dependence of the background
fields, so we assume the metric and the tensor field to be independent
of $X^+$. If we assume, in addition, that they are independent of
$X^-$, it turns out that $P_-$ becomes $\tau$-independent. This
allows us 
to set $P_-(\s)=(P_-)_0\,\sqrt{w(\s)}$, exactly as in flat space. 
With these restrictions,
it is possible to write down a gauge theory of area-preserving 
diffeomorphisms for the membrane in the presence of background 
fields. Its Lagrangian density equals  
\bea
{w}^{-1/2}\, {\cal L} &=& \ft{1}{2} (D_0 X^a)^2 + D_0 X^a \left(
\ft{1}{2} C_{abc} \{ X^b, X^c \} + G_{a+} \right) \nonumber \\
&&- \ft{1}{4}\{ X^a, X^b \}^2 + \ft12{G_{++}} + \ft{1}{2}
C_{+ab} \{ X^a, X^b \} \, , \label{BGlagr}
\eea
where we used the metric $G_{ab}$ to contract transverse
indices; the Poisson bracket and the covariant derivatives were 
already introduced in section~2. For convenience we have set 
$(P_-)_0 =1$.  

The action corresponding to \eqn{BGlagr} is manifestly invariant 
under area-preserving diffeomorphisms in the presence of the 
background fields. It  
is now straightforward to write it in terms of a matrix model, 
by truncating the mode expansion for coordinates and momenta 
as explained in the previous section.  
Matrix models in curved space have been discussed before 
\cite{oldBG}; for more recent papers dealing with matrix models in 
the presence of certain backgrounds, see \cite{newBG}. A more 
explicit derivation of the results of this section and their 
supersymmetric extension will appear in a forthcoming 
publication~\cite{DWPP4}.        
                
%%%%%%%%%%%%%%%%%%%%%%%%%%%%%%%%%%%%%%%%%%%%%%%%%%%%%%%%%%% 
\section{The continuous supermembrane mass spectrum}

The continuous mass spectrum of the supermembrane
forms an obstacle in interpretating the membrane states as elementary 
particles, in analogy to what is done in string theory. 
Instead the continuity of the spectrum should be viewed as a  
result of the fact that  
supermembrane states do not really exist as asymptotic states. The 
membrane collapses into stringlike configurations and is to be 
interpreted as a multimembrane state. Obviously such states 
exhibit a continuous mass spectrum. As we alluded to earlier, 
there is evidence that massless ground states exist, probably 
associated with the states of 11-dimensional  
supergravity \cite{mboundstates}. In the winding sector there may exist  
massive BPS states, which are the lowest-mass states for given 
winding number. Whether additional non-BPS bound states exist is not 
known. It could be that beyond the massless and BPS winding 
states, there is nothing than a continuum of multimembrane states. 

Qualitatively, the situation is the same for the matrix models 
\eqn{hamiltonian} based on a finite number of degrees of freedom. 
Among the zero-energy states there are those where the matrices take a 
block-diagonal form, which can be regarded as a direct product of 
states belonging to lower-rank matrix models \cite{BFSS}. The 
fact that the moduli space of ground states, whose nature is 
protected by supersymmetry at the quantum-mechanical level, is 
isomorphic to ${\bf R}^{9N}/S_N$, is already indicative of a 
corresponding description in terms of an $N$-particle Fock space. 
The finite-$N$ matrix models have an independent interpretation 
in string theory. Strings can end on certain defects by means of  
Dirichlet boundary conditions. These defects are 
called D-branes (for further references, see \cite{Dbranes}). 
They can have a $p$-dimensional spatial extension 
and carry Ramond-Ramond charges \cite{Polchinski}. 
D-Branes play an important role in the nonperturbative 
behaviour of string theory. The models of section~1 are relevant 
for D0-branes (Dirichlet particles), but we note in passing 
that there are similar models relevant for higher-dimensional 
D-branes, which emerge in the zero-volume limit of supersymmetric 
gauge theories coupled to matter. 

The effective short-distance description for D-branes can be 
derived from simple arguments \cite{boundst}. As the strings must be 
attached to the  $p$-dimensional branes, we are dealing with open 
strings whose endpoints are attached to a  
$p$-dimensional subspace. At short distances, the interactions 
caused by these open strings are determined by the massless 
states of the open string, which constitute the ten-dimensional 
Yang-Mills supermultiplet, propagating in a reduced ($p+1$)-dimensional 
spacetime. Because the endpoints of open strings carry Chan-Paton 
factors the effective  
short-distance behaviour of $N$ D-branes can be described in terms 
of a U($N$) ten-dimensional supersymmetric gauge theory reduced 
to the $(p+1$)-dimensional world volume of the D-brane. The 
U(1) subgroup is associated with the center-of-mass motion of the $N$ 
D-branes.                                                     

In the type-IIA superstring one has Dirichlet 
particles moving in a 9-dimensional space. As the world volume 
of the particles is 
one-dimensional ($p=0$), the short-distance interactions between 
these particle is thus described by the model of section~1 with 
gauge group U($N$) and $d=9$. The continuous spectrum without 
gap is natural here, as it is known that, for static D-branes, the 
Ramond-Ramond repulsion cancels against the gravitational and 
dilaton atraction, a similar phenomenon as for BPS 
monopoles. With this gauge group the  
coordinates can be described in terms of $N\times N$ hermitean 
matrices. The valley configurations correspond to the situation 
where all these matrices can be diagonalized simultanously. The 
eigenvalues then define the positions of $N$  
D-particles in the 9-dimensional space. As soon as one or several 
of these particles coincide   
then the [U(1)]$^N$ symmetry that is left invariant in the 
valley, will be enhanced to a nonabelian 
subgroup of U($N$). Clearly there are more degrees of freedom 
than those corresponding to the D-particles, which are associated 
with the strings stretching between 
the D-particles. As we alluded to above the model naturally 
incorporates configurations corresponding to widely separated 
clusters of D-particles, each of which can be described by a 
supersymmetric quantum-mechanics model based on the product of a 
number of U($k$) subgroups forming a maximal commuting subgroup 
of U($N$). When all the D-particles move further apart this 
corresponds to configurations deeper and deeper into the 
potential valleys.  
These D-particles thus define an independent perspective on the 
models introduced in section~1, which can 
be used to study their dynamics. We refer to \cite{Dpart} for  
work along these lines. 

The study of D-branes was further motivated by a conjecture 
according to which the degrees of freedom of M-theory 
are fully captured by the U$(N)$ super-matrix models in the $N\to 
\infty$ limit \cite{BFSS}. The elusive M-theory is defined as the 
strong-coupling limit of type-IIA   
string theory and is supposed to capture all the relevant degrees 
of freedom of all known string theories, both at the perturbative 
and the nonperturbative level \cite{Townsend,witten3}. In 
this description the various string-string dualities are fully 
incorporated. At large distances M-theory is described by 
11-dimensional supergravity. A direct relation between 
supermembranes and type-IIA string theory was emphasized in 
\cite{Townsend}, based on the relation between extremal
black holes in 10-dimensional supergravity \cite{HorStrom} and 
the Kaluza-Klein states of 11-dimensional supergravity 
in an $S^1$ compactification. In this compactification the 
Kaluza-Klein photon coincides with the Ramond-Ramond vector field of 
type-IIA string theory. Therefore Kaluza-Klein states are 
BPS states whose Ramond-Ramond charge is proportional to their 
mass. Hence they have the same characteristics as the Dirichlet 
particles. On the other hand, the effective interaction between 
infinitely many Dirichlet particles leads to a theory that is 
identical to that of an elementary supermembrane. There are 
alternative compactifications of  
M-theory which make contact with other string theories. 
Supermembranes have been used to provide evidence for the duality 
of M-theory on $\mbox{\bf R}^{10}\times S_1/\mbox{\bf Z}_2$ and 
10-dimensional $E_8\times E_8$ heterotic strings 
\cite{horvw}. Finally the so-called double-dimensional 
reduction of membranes leads to fundamental string 
states \cite{DHIS}.
                                                          
\vspace{.5cm}

%%%%%%%%%%%%%%%%%%%%%%%%%%%%%%%%%%%%%%%%%%%%%%%%%%%%%%%%%
\noindent{\bf Acknowledgements}\\[1ex]
Most of the work reported here was carried out in collaboration 
with K. Peeters and J.C. Plefka. I thank the organizers of this 
symposium for providing a stimulating atmosphere.
%%%%%%%%%%%%%%%%%%%%%%%%%%%%%%%%%%%%%%%%%%%%%%%%%%%%%%%%%%

\end{document}